\documentclass[a4paper, 12pt]{article}

\usepackage{amsmath}
\usepackage{amssymb}
\usepackage{graphics}

\begin{document}
\renewcommand{\theequation}{\thesection .\arabic{equation}}

\newcommand{\sign}{\operatorname{sign}}
\newcommand{\Ci}{\operatorname{Ci}}
\newcommand{\tr}{\operatorname{tr}}

\newcommand{\beq}{\begin{equation}}
\newcommand{\eeq}{\end{equation}}
\newcommand{\beqn}{\begin{eqnarray}}
\newcommand{\eeqn}{\end{eqnarray}}

\newcommand{\slp}{\raise.15ex\hbox{$/$}\kern-.57em\hbox{$ \partial $}}
\newcommand{\lnA}{\raise.15ex\hbox{$/$}\kern-.57em\hbox{$A$}}
\newcommand{\unmedio}{{\scriptstyle\frac{1}{2}}}
\newcommand{\uncuarto}{{\scriptstyle\frac{1}{4}}}

\newcommand{\trial}{_{\text{trial}}}
\newcommand{\true}{_{\text{true}}}
\newcommand{\const}{\text{const}}

\newcommand{\intp}{\int\frac{d^2p}{(2\pi)^2}\,}
\newcommand{\intx}{\int d^2x\,}
\newcommand{\inty}{\int d^2y\,}
\newcommand{\intxy}{\int d^2x\,d^2y\,}

\newcommand{\bP}{\bar{\Psi}}
\newcommand{\bc}{\bar{\chi}}
\newcommand{\hs}{\hspace*{0.6cm}}

\newcommand{\bra}{\left\langle}
\newcommand{\ket}{\right\rangle}
\newcommand{\bracket}{\left\langle\,\right\rangle}

\newcommand{\D}{\mbox{$\mathcal{D}$}}
\newcommand{\N}{\mbox{$\mathcal{N}$}}
\newcommand{\Lag}{\mbox{$\mathcal{L}$}}
\newcommand{\V}{\mbox{$\mathcal{V}$}}
\newcommand{\Z}{\mbox{$\mathcal{Z}$}}

\begin{titlepage}


\vspace{2cm}

\begin{center}

{\Large {\bf Dynamics of phantom  model with O(N) symmetry in loop
quantum cosmology}}

\vspace{1.3cm}

Zu-Yao Sun$^{a}$\footnote{e-mail: zysun@shmtu.edu.cn} ,Chun-Xiao
Yue$^{b}$ , You-Gen Shen$^{c}$\footnote{e-mail: ygshen@shao.ac.cn},
Chang-Bo Sun$^{d}$

\vspace{.8cm}

$^a$ {\it College of Arts and Sciences, Shanghai Maritime
University, Shanghai 200135, China.}

\smallskip

$^b$ {\it Shanghai Jianqiao College, Shanghai 201319, China.}

\smallskip

$^c$ {\it Key Laboratory for Research in Galaxies and Cosmology,
Shanghai Astronomical Observatory, Chinese Academy of Sciences, 80
Nandan RD, Shanghai 200030, China }

\smallskip
$^d$ {\it College of science, Shaanxi university of science and
tecnology, Xi'an, Shanxi 710021, China}

\vspace{1cm}

\begin{abstract}
Many astrophysical data show that the expansion of our universe is
accelerating. In this paper, we study the model of phantom with O(N)
symmetry in background of loop quantum cosmology(LQC). We
investigate the phase-space stability of the corresponding
autonomous system and find no stable node but only 2 saddle points
in the field of real numbers. The dynamics is similar to the
single-field phantom model in LQC\cite{lqc1}. The effect of  O$(N)$
symmetry just influence the detail of the universe's evolution. This
is a sharp contrast with the result in general relativity, in which
the dynamics of scalar fields models with O$(N)$ symmetry are quite
different from the single-field
models\cite{onquintessence},\cite{onphantom},\cite{onquintom}.
 \end{abstract}
\end{center}

\end{titlepage}

\newpage
\section{Introduction}\

Recently, many astrophysical data, such as WMAP\cite{cmb}, type Ia
supernovae\cite{sia} and large scale galaxy surveys\cite{lsgs}, show
that our universe is almost flat, and currently undergoing a period
of accelerating expansion which is driven by a yet unknown dark
energy. About $70\%$ of the energy density in our universe is dark
energy nowadays. There are many candidates of dark energy such as
the cosmological constant\cite{cc},
quintessence\cite{q1,q2,q3,q4,q5,onquintessence},
phantom\cite{p1,p2,p3,onphantom}, quintom\cite{quintom,onquintom}
etc. The equation-of-state parameter $w=p/\rho$ plays an important
role in these models. The parameter of quintessence and phantom lies
in the range $-1/3>w>-1$ and $w<-1$ respectively. And for quintom
$w$ could cross -1.

Li $et$ $al.$ generalized the quintessence and phantom models to
fields possess O$(N)$ symmetry\cite{onquintessence,onphantom}.
Setare $et$ $al.$ extended the generalization to the quintom
model\cite{onquintom}. In the above papers they showed that the
behaviors of the dynamic system with O$(N)$ symmetry are quite
different from the single-field models.

Many works on dark energy have been done in the framework of
Einstein's classical general relativity(GR). However, most
physicists believe that the gravity should be quantized in a
ultimate theory. Thus, we should consider the quantum effects in the
evolution of our universe. Loop quantum gravity(LQG) is a
outstanding candidate of quantum gravity theory. LQG is a background
independent and non-perturbative theory. At the quantum level, the
classical spacetime continuum is replaced by a discrete quantum
geometry and the operators corresponding to geometrical quantities
have discrete eigenvalues. Many topics on cosmology are discussed
within the framework of LQG in various literature in which it is
known as Loop Quantum Cosmology
(LQC)\cite{lqc1,lqc2,lqc3,lqc4,lqc5}. Recently, Samart $et$
$al.$\cite{lqc1} have studied the phantom dynamics in LQC. The work
of quintom and hessence model in loop quantum cosmology have been
done by Wei $et$ $al.$\cite{lqc3}. They found the behaviors is
different
 from the ones in standard FRW, such as the avoidance of
future singularities.

In this paper, we investigate the evolution of universe dominated by
phantom with O(\textit{N}) symmetry. The phase-space stability of
the corresponding autonomous system has been discussed. We find
there's no stable attractor but only 2 saddle points, this result is
similar to the single-field phantom in LQC. In order to research the
evolution of our model, we numerically study the system in
exponential potential, and draw 2 graph to display the evolution of
Hubble parameter $H$ and energy density $\rho$ to cosmic time $t$.
The figures reveal that in LQC even for a universe dominated by
phantom, the Hubble parameter $H$ may turn negative and oscillate.
Thus, the universe will enter a oscillatory regime. The relation
between $\rho$ and $H$ is also shown in a graph.
\section{Loop quantum cosmology}\

Loop quantum cosmology(for review, see Ref.\cite{lqcre1,lqcre2}) can
describe the homogeneous and isotropic spacetimes by effective
modified Friedmann equation for flat universe. The quantum effects
could be reflected by adding a correction term $\rho/\rho_{c}$ into
the eqution. In Einstein's general relativity, the standard
Friedmann equation which describes a flat universe is:
\begin{equation}\label{stfre}
H^{2}=\frac{\kappa^{2}}{3}\rho
\end{equation}
Instead of Eq. (\ref{stfre}), the modified Friedmann equation for a
flat universe in LQC is given by \cite{mfe1,lqc2}
\begin{equation}\label{mdfre}
H^{2}=\frac{\kappa^{2}}{3}\rho(1-\frac{\rho}{\rho_{c}})
\end{equation}
As in GR, $H=\dot{a}/a$ is Hubble parameter, $\rho$ is the total
energy density, and a dot denotes the derivative with respect to
cosmic time t. $\rho_{c}$ is critical loop quantum density:
\begin{equation}\label{rhoc}
 \rho_c=\frac{\sqrt{3}}{16\pi^2\gamma^3 G^2 \hbar},
\end{equation}
Where $\gamma$ is the dimensionless Barbero-Immirzi parameter and is
suggested that $\gamma\simeq 0.2375$ by the black hole
thermodynamics in LQG\cite{gama}. The modified Friedmann equation
provides an effective description for LQC which very well
approximates the underlying discrete quantum dynamics.
Differentiating Eq. (\ref{mdfre}) and use the conservation law
\begin{equation}\label{cl}
\dot{\rho}+3H\left(\rho+p\right)=0
\end{equation}
We obtain the effective modified Raychaudhuri equation
\begin{equation}\label{ray}
\dot{H}=-\frac{\kappa^{2}}{2}\left(\rho+p\right)(1-2\frac{\rho}{\rho_{c}})
\end{equation}

where $p$  is the total pressure.  In LQC the accelerating condition
is
\begin{equation}\label{ac}
\frac{\ddot{a}}{a}=\dot{H}+H^{2}>0
\end{equation}
The dynamics of our universe could be analyzed through
Eq.(\ref{mdfre}), (\ref{cl}), (\ref{ray}) and the equation of the
fields which dominate the universe.
\section{O(N) Phantom}\

We consider the flat Robertson-Walker metric
\begin{equation}\label{metric}
ds^{2}=dt^{2}-a^{2}(t)\left(dx^{2}+dy^2+dz^{2}\right)
\end{equation}

The Lagrangian density for the Phantom with
 O(\textit{N}) symmetry is given by\cite{onphantom}

\begin{equation}\label{lagrangian}
L_{\Phi}=-\frac{1}{2}g^{\mu\nu}(\partial_{\mu}\Phi^{a})(\partial_{\nu}\Phi^{a})-V(|\Phi^{a}|)
\end{equation}

\noindent where $\Phi^{a}$ is the component of the scalar field,
$a=1,2,\cdots,N$. In order to impose the O(\textit{N}) symmetries,
following\cite{onphantom}, we write it in the form

\begin{eqnarray}\label{imag}
\Phi^{1}=R(t)\cos\varphi_{1}(t)\hspace{4.2cm}\nonumber\\
\Phi^{2}=R(t)\sin\varphi_{1}(t)\cos\varphi_{2}(t)\hspace{2.85cm}\nonumber\\
\Phi^{3}=R(t)\sin\varphi_{1}(t)\sin\varphi_{2}(t)\cos\varphi_{3}(t)\hspace{1.5cm}\\
\cdots\cdots\hspace{4cm}\nonumber\\
\Phi^{N-1}=R(t)\sin\varphi_{1}(t)\cdots\sin\varphi_{N-2}(t)\cos\varphi_{N-1}(t)\hspace{-0.35cm}\nonumber\\
\Phi^{N}=R(t)\sin\varphi_{1}(t)\cdots\sin\varphi_{N-2}(t)\sin\varphi_{N-1}(t)\nonumber
\end{eqnarray}

Thus, we have $|\Phi^{a}|=R$ and assume that the potential
V($|\Phi^{a}|$) depends only on \textit{R}. The radial equation of
scalar fields is
\begin{eqnarray}\label{kg}
\ddot{R}&+&3H\dot{R}-\frac{\Omega^{2}}{a^{6}R^{3}}-\frac{\partial
V(R)}{\partial R}=0
\end{eqnarray}
The ``angular components'' contribute a effective term
$\frac{\Omega^{2}}{a^{6}R^{3}}$ to the system's dynamics. The energy
density $\rho$ and the pressure $p$ of the O(\textit{N}) phantom are
given as
\begin{equation}\label{rho}
\rho=-\frac{1}{2}(\dot{R}^{2}+\frac{\Omega^{2}}{a^{6}R^{2}})+V(R)
\end{equation}
and
\begin{equation}\label{p}
p=-\frac{1}{2}(\dot{R}^{2}+\frac{\Omega^{2}}{a^{6}R^{2}})-V(R)
\end{equation}
Hence, the equation-of-state for the O(\textit{N}) phantom is
\begin{equation}\label{w}
w=\frac{-\frac{1}{2}(\dot{R}^{2}+\frac{\Omega^{2}}{a^{6}R^{2}})-V(R)}{-\frac{1}{2}(\dot{R}^{2}+\frac{\Omega^{2}}{a^{6}R^{2}})+V(R)}
\end{equation}
Different from GR, the accelerating condition for universe in LQC is
not $w<-1/3$ but Eq.(\ref{ac}).

\section{EVOLUTION AND STABILITY OF THE MODEL}\

 In this section we discuss the dynamics of universe dominated by phantom with
O(\textit{N})
 symmetry in LQC. In order to get a possible evolution of the
model, we choose exponential potential
\begin{equation}\label{v}
V=V_{0}e^{-\lambda\kappa R},
\end{equation} The evolution of universe is governed by
 Eq.(\ref{mdfre}),(\ref{ray}),(\ref{kg}).  To investigate the stability
 of the model, we
 introduce the following dimensionless variables
\begin{eqnarray}\label{newvar}
x&=&\frac{\kappa\dot{R}}{\sqrt{6}H},\
y=\frac{\kappa\sqrt{V(R)}}{\sqrt{3}H},\
z=\frac{\kappa}{\sqrt{6}H}\frac{\Omega}{a^3R},
\\
m&=&\frac{\rho}{\rho_{c}},\ \xi=\frac{1}{\kappa R},\ \ N=\ln a
\end{eqnarray}
Using these variables, Eq.(\ref{mdfre}),(\ref{ray}),(\ref{kg}) can
be written as the following autonomous system:
\begin{eqnarray}\label{auto1}
\nonumber \frac{dx}{dN}&=&-3x+\sqrt{6}\xi
z^{2}-\sqrt{\frac{3}{2}}\lambda y^{2}-(3x^{3}+3xz^{2})(1-2m)
\\ \nonumber
\frac{dy}{dN}&=&-\sqrt{\frac{3}{2}}xy\lambda-(3x^{2}y+3yz^{2})(1-2m)
\\
\frac{dz}{dN}&=&-3z-\sqrt{6}xz\xi-(3x^{2}z+3z^{3})(1-2m)
\\ \nonumber
\frac{dm}{dN}&=&-3m(1+\frac{-x^{2}-y^{2}-z^{2}}{-x^{2}+y^{2}-z^{2}})\\
\nonumber
\frac{d\xi}{dN}&=&-\sqrt{6}x\xi^{2} \nonumber
\end{eqnarray}

In the following, it is straightforward to analyze the critical
points as well as their stability. We can get critical points by
setting the right hand of the equation(\ref{auto1}) to zero. As
presented in table \ref{critical}, in the field of real numbers
there are only 2 real critical points: $A$ and $B$. We expand the
variables around the critical point $A$ and $B$ in the form
$x=x_{c}+\delta x, y=y_{c}+\delta y, z=z_{c}+\delta z,
m=m_{c}+\delta m, \xi=\xi_{c}+\delta \xi$, where $\delta x, \delta
y, \delta z, \delta m, \delta \xi$ are perturbations of the
variables near the critical points and we consider them forming a
column vector denoted as U. Substituting the above expression into
(\ref{auto1}) and neglecting higher order terms in the
perturbations, we can obtain the equations for the perturbations up
to first order as:
\begin{equation}\label{u}
\textbf{U}'={\bf{M}}\cdot \textbf{U},
\end{equation}
where the prime denotes differentiation with respect to N. M is a
$5\times5$ matrix formed by the coefficients of the perturbation
equations (\ref{u}). The matrix for the critical point ($x_{c},
y_{c}, z_{c}, m_{c}, \xi_{c}$) is given by
\begin{eqnarray}
 \label{matrix}
M=\left( \begin{array}{ccccc}
\frac{\partial x'}{\partial x}& \frac{\partial x'}{\partial y}& \frac{\partial x'}{\partial z}& \frac{\partial x'}{\partial m}& \frac{\partial x'}{\partial \xi}\\
\frac{\partial y'}{\partial x}& \frac{\partial y'}{\partial y}& \frac{\partial y'}{\partial z}& \frac{\partial y'}{\partial m}& \frac{\partial y'}{\partial \xi}\\
\frac{\partial z'}{\partial x}& \frac{\partial z'}{\partial y}& \frac{\partial z'}{\partial z}& \frac{\partial z'}{\partial m}& \frac{\partial z'}{\partial \xi}\\
\frac{\partial m'}{\partial x}& \frac{\partial m'}{\partial y}& \frac{\partial m'}{\partial z}& \frac{\partial m'}{\partial m}& \frac{\partial m'}{\partial \xi}\\
\frac{\partial \xi'}{\partial x}& \frac{\partial \xi'}{\partial y}&
\frac{\partial \xi'}{\partial z}& \frac{\partial \xi'}{\partial m}&
\frac{\partial \xi'}{\partial \xi}
\end{array} \right)_{(x=x_{c}, y=y_{c}, z=z_{c}, m=m_{c}, \xi=\xi_{c})}\,.
\end{eqnarray}
The eigenvalue of (\ref{matrix}) determine the type and stability of
the critical points. We present them in table \ref{eigen}. As shown,
both critical A and B are saddle points, it follows immediately that
the phase trajectory is very sensitive to initial conditions given
to the system. This result is similar to the single-field phantom in
LQC\cite{lqc1}. It is interesting to contrast the influence of
O$(N)$ symmetry in GR and LQC. Several
works\cite{onquintessence}\cite{onphantom}\cite{onquintom} shows
that the dynamics of quintessence, phantom and quintom with O$(N)$
symmetry in GR is quite different from the single-field model.
\begin{center}
\begin{table}
\begin{tabular}{| c| c| c |c| c |c |c |}
  \hline
  Critical point &   $x_{c}$ & $y_{c}$ & $z_{c}$ & $m_{c}$ & $\xi_{c}$ \\
\hline
   A &   $-\frac{\lambda}{\sqrt{6}}$ & $-\frac{\sqrt{6+\lambda^{2}}}{\sqrt{6}}$ & 0 & 0 & 0 \\
\hline
  B &   $-\frac{\lambda}{\sqrt{6}}$ & $\frac{\sqrt{6+\lambda^{2}}}{\sqrt{6}}$ & 0 & 0 & 0\\
 \hline
\end{tabular}
\caption[crit]{\label{critical} The real critical points of the
autonomous system (\ref{auto1}).}
\end{table}
\end{center}
\begin{center}
\begin{table}
\begin{tabular}{ |c| c |c |c |c |c |}
  \hline
  Label & Eigenvalues & Stability & $w$ & Acceleration\\
\hline
A&$-\lambda^{2}$ , $\frac{1}{2}(-6-\lambda^{2})$ , $\frac{1}{2}(-6-\lambda^{2})$ , $\lambda^{2}$, 0& saddle point& $-1-\frac{\lambda^{2}}{3}$ & Yes\\
\hline
B&$-\lambda^{2}$ , $\frac{1}{2}(-6-\lambda^{2})$ , $\frac{1}{2}(-6-\lambda^{2})$ , $\lambda^{2}$, 0& saddle point& $-1-\frac{\lambda^{2}}{3}$ & Yes\\
 \hline
\end{tabular}
\caption[crit]{\label{eigen} The eigenvalues of the matrix M and the
properties of critical points in the autonomous system
(\ref{auto1}).}
\end{table}
\end{center}

\begin{figure}[htbp]
\centerline{\includegraphics{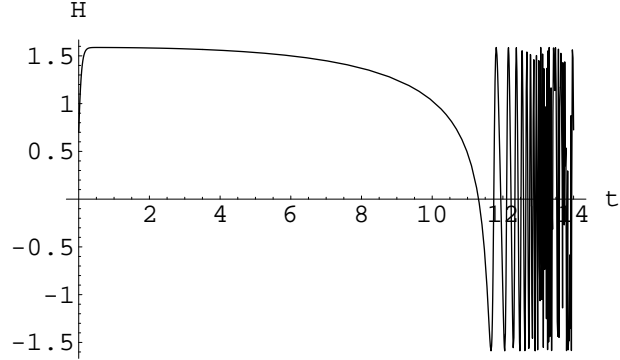}} \caption{The evolution of $H$
with respect to $t$ for $\lambda=-0.1$, $\Omega=1$,
$\kappa=\sqrt{8\pi}$, $\rho_{c}=1.5$, $V_{0}=1.2$.} \label{ht}
\end{figure}

\begin{figure}[htbp]
\centerline{\includegraphics{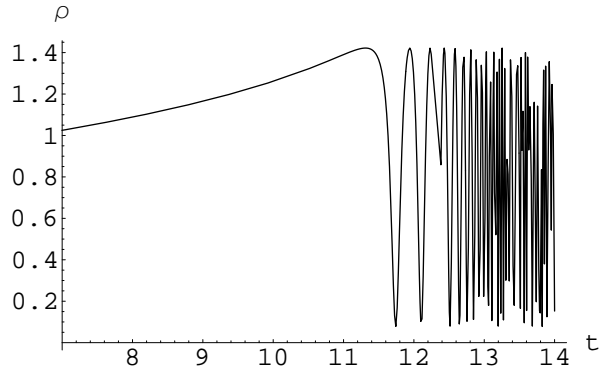}} \caption{The evolution of
cosmic total energy density $\rho$ with respect to $t$ for
$\lambda=-0.1$, $\Omega=1$, $\kappa=\sqrt{8\pi}$, $\rho_{c}=1.5$,
$V_{0}=1.2$.} \label{rhot}
\end{figure}

\begin{figure}[htbp]
\centerline{\includegraphics{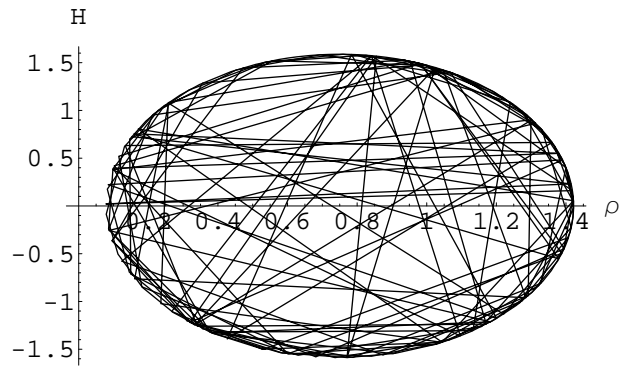}} \caption{The relation
between $H$ and $\rho$ while t goes from $0$ to $14$. Set values are
$\lambda=-0.1$, $\Omega=1$, $\kappa=\sqrt{8\pi}$, $\rho_{c}=1.5$,
$V_{0}=1.2$.} \label{hrho}
\end{figure}

In the following, we study the possible evolution of our model
numerically. In order to do this, we set the value of some
parameters: $\lambda=-0.1$, $\Omega=1$, $\kappa=\sqrt{8\pi}$,
$\rho_{c}=1.5$, $V_{0}=1.2$, and draw 3 graphs: Fig.\ref{ht},
Fig.\ref{rhot}, Fig.\ref{hrho}. Fig.\ref{ht} shows the evolution of
Hubble parameter $H$ with respect to cosmic time $t$. From
Fig.\ref{ht}, we found $H$ may take maximum value while
$\rho=\rho_{c}/2$, and then decline even to negative interval. After
$H$ taking minimum value, it would bounce and then oscillate. Thus,
the accelerating expansion of our universe would end in  finite time
and undergoes  contraction from halting($H=0$). Then the universe
enters an oscillating stage. Fig.\ref{rhot} shows the evolution of
cosmic total energy density $\rho$, it oscillate while $t$  goes by.
In fact, from Fig.\ref{hrho} we can get some insight into the
evolution of the model. Fig.\ref{hrho} reveals the relation between
$H$ and $\rho$. As time goes by, the trajectory makes a closed
region, and the oscillation of $H$ and $\rho$ are restricted in a
certain interval.

\section{Conclusions}\

In this paper, we investigate the dynamics of phantom model with
O$(N)$ symmetry in background of loop quantum cosmology. We discuss
its stability by analyzing the autonomous system Eq.(\ref{auto1}).
It shows that there is no attractor in the system, but only 2 saddle
points. This result is similar to the single-field model in
LQC(\cite{lqc1}). The effect of  O$(N)$ symmetry just influence the
detail of the universe's evolution. It is a sharp contrast with the
result in GR, in which the dynamics of scalar fields models with
O$(N)$ symmetry are quite different from the single-field models.
Finally, to show some insight into our model, we draw 3 graphs by
Numerical analysis. The possible oscillation of $H$ and $\rho$  are
revealed in Fig.\ref{ht}. and Fig.\ref{rhot}. Fig.\ref{hrho} shows
that as time goes by, $H$ and $\rho$ are restricted in a certain
interval, and the trajectory makes a closed region.
\section*{Acknowledgements}

\hspace*{7.5mm} The work has been supported by the National Natural
Science Foundation of China (Grant No. 10973014) and special fund for academic development
of young teachers in shanghai colleges and universities.  (Grant No. shs-07025.).\\


\begin{thebibliography}{99}
\bibitem{lqc1} Daris Samart and Burin Gumjudpai, Phys.Rev.D76:043514,2007.
arXiv:0704.3414.
\bibitem{onquintessence}X. Zh. Li, J. G. Hao and D. J. Liu, Class. Quant. Grav. {\bf{19}}, 6049
(2002) [arXiv:astro-ph/0107171].
\bibitem{onphantom}X. Li and J. Hao, Phys. Rev. D {\bf{69}}, 107303 (2004) [arXiv:hep-th/0303093].
\bibitem{onquintom} M. R. Setare and E. N. Saridakis, JCAP
0809:026,2008 [arXiv:0809.0114]
\bibitem{cmb}E. Komatsu et al.[WMAP Collaboration], Astrophys. J. Suppl. 180, 330 (2009) [arXiv:0803.0547 [astro-ph]].
\bibitem{sia}M. Hicken et al.,  Astrophys. J. 700, 1097 (2009) [arXiv:0901.4804 [astro-ph.CO]].
\bibitem{lsgs}R. Scranton et al. [SDSS Collaboration], [arXiv:astroph/ 0307335].
\bibitem{cc}P. J. Peebles and B. Ratra, Rev. Mod. Phys. 75, 559 (2003)
[arXiv:astro-ph/0207347].
\bibitem{q1}Caidwell.R.R, Dave.R and Steinhardt.P.J, Phys.Rev.Lett.80 (1998)
1582
\bibitem{q2}C. J. Gao and Y. G. Shen.(2002). Phys.Lett.{\bf B541},1
\bibitem{q3}C. J. Gao and Y. G. Shen.(2002). Chin.Phys.Lett.{\bf 19},1396
\bibitem{q4}P. J. E. Peebles, B. Ratra, Rev. Mod. Phys. 75(2003)599 ; T. Padmanabhan, Phys. Rept.
380(2003)235
\bibitem{q5}R. R. Caldwell, R. Dave and P. J. Steinhardt, Phys. Rev. Lett. 80(1998)1582
\bibitem{p1}J.-An Gu and W. -Y. P. Hwang.(2001) Phys.Lett. {\bf B517}, 1
\bibitem{p2}Caidwell.R.R, Phys.Lett. B 545 (2002) 23.
\bibitem{p3}Z. Y. Sun and Y. G. Shen. Gen.Rel.Grav. 37(2005) 243
\bibitem{quintom}B. Feng, X. L. Wang and X. M. Zhang, Phys. Lett. B 607, 35 (2005)
[astro-ph/0404224].
\bibitem{lqc2}Xin Zhang and Yi Ling, JCAP0708:012,2007. [arXiv:0705.2656]
\bibitem{lqc3} Hao Wei and Shuang Nan Zhang,
Phys.Rev.D76:063005,2007. [arXiv:0705.4002]
\bibitem{lqc4}Xiangyun Fu, Hongwei Yu and Puxun Wu,
PhysRevD.78.063001. [arXiv:0808.1382]
\bibitem{lqc5} Songbai Chen, Bin Wang and Jiliang Jing, arXiv:0808.3482
\bibitem{lqcre1} M. Bojowald, Living Rev. Rel. 8, 11 (2005)
    [gr-qc/0601085];\\ M. Bojowald, gr-qc/0505057.
\bibitem{lqcre2} A. Ashtekar, M. Bojowald and J.
    Lewandowski, Adv. Theor. Math. Phys. 7, 233 (2003)
\bibitem{mfe1}A. Ashtekar, AIP Conf. Proc. 861, 3 (2006) [gr-qc/0605011].
[gr-qc/0304074];\\ A. Ashtekar, gr-qc/0702030.
\bibitem{gama}A. Ashtekar, J. Baez, A. Corichi and K. Krasnov, Phys. Rev. Lett. 80, 904 (1998) [gr-qc/9710007];
M. Domagala and J. Lewandowski, Class. Quant. Grav. 21, 5233 (2004)
[gr-qc/0407051]; K. A. Meissner, Class. Quant. Grav. 21, 5245 (2004)
[gr-qc/0407052].

\bibitem{h1}A. Cohen, D. Kaplan and A. Nelson, hep-th/9803132, Phys. Rev. Lett.
82 (1999) 4971.
\bibitem{h2}S.D.H. Hsu, hep-th/0403052.
\bibitem{h3}M. Li, hep-th/0403127
\bibitem{h4}Yungui Gong, Phys.Rev.D 70(2004) 064029
\bibitem{in1}A.Zee,Phys.Rev.Lett. 42(1979)417
\bibitem{in2}Frank S.Accetta, David J.Zoller and Michael S.Turner,
Phys.Rev.D 31(1985) 3046
\bibitem{in3}Y. G. Shen Chin. Phys. Lett 12(1995) 509
\bibitem{h5}W. Fischler and L. Susskind, hep-th/9806039; R. Bousso, JHEP 9907 (1999) 004.
\end{thebibliography}
\end{document}